\documentclass[%
 reprint,
superscriptaddress,
nofootinbib,
 amsmath,amssymb,
 aps,
floatfix
]{revtex4-2}

\usepackage{siunitx}
\usepackage[english]{babel}
\usepackage{tabulary}
\usepackage[utf8]{inputenc} 
\usepackage{xspace}
\usepackage{dcolumn}
\usepackage{bm}
\usepackage{hyperref}
\usepackage{paralist}
\usepackage{booktabs}
\usepackage{tikz,pgf,epsfig,wasysym,mathrsfs}
\usetikzlibrary{quantikz}
\usepackage{xcolor}
\usepackage{siunitx}
\usepackage{multirow}
\usepackage{overpic}
\usepackage{amsmath}
\usepackage{braket}
\usepackage[ruled,linesnumbered]{algorithm2e}
\usepackage{graphicx}
\usepackage{dcolumn}
\usepackage{bm}
\usepackage{multirow}
\usepackage{amsmath}
\usepackage{braket}
\usepackage{graphicx}
\usepackage{dcolumn}
\usepackage{bm}

\newcommand{\probdefbis}[6][Task]{
\hbox{\vbox{
\begin{quote}
  \label{#6}
  \ifthenelse{\equal{#4}{}}{}{{#4}\ifthenelse{\equal{#5}{}}{}{({#5})}}
  \begin{compactdesc}
    \item [Input]{#2}
    \item [#1]{#3}
  \end{compactdesc}
\end{quote}
}}
}
\newcommand{\SSSAT}{\textsc{3-Sat}\xspace}
\newcommand{\MAXCLIQUE}{\textsc{Maximum Clique}\xspace}
\newcommand{\MAXSSAT}{\textsc{Maximum 2-Sat}\xspace}
\newcommand{\MAXSSSAT}{\textsc{Maximum 3-Sat}\xspace}
\newcommand{\MAXCUT}{\textsc{Maximum Cut}\xspace}
\newcommand{\MWIS}{\textsc{Maximum Weighted Independent Set}\xspace}
\newcommand{\PARTITION}{\textsc{Minimum Partition}\xspace}

\ifx\newdefinition\undefined
  \let\newdefinition\newtheorem
\fi
\ifx\definition\undefined
  \newdefinition{definition}{Definition}

\fi

\newdefinition{construction}{Construction}

\begin{document}

\preprint{APS/123-QED}
\title{
Solving various NP-Hard problems using exponentially 
 fewer qubits\\ on a Quantum Computer}

\author{Yagnik Chatterjee}
\email{yagnik.chatterjee@totalenergies.com}
\affiliation{TotalEnergies,
Tour Coupole - 2 place Jean Millier 92078 Paris la Défense cedex, France}
\affiliation{LIRMM, Université de Montpellier, CNRS, 161 rue Ada, 34392 Montpellier Cedex 5}

\author{Eric Bourreau}
\email{eric.bourreau@lirmm.fr}
\affiliation{LIRMM, Université de Montpellier, CNRS, 161 rue Ada, 34392 Montpellier Cedex 5}
\author{Marko J. Ran\v{c}i\'{c}}
\email{current address: University of Heidelberg, Institute of Computer Engineering, Im Neuenheimer Feld 368, 69120 Heidelberg, Germany. email: marko.rancic@ziti-uni-heidelberg.de}
\affiliation{TotalEnergies,
Tour Coupole - 2 place Jean Millier 92078 Paris la Défense cedex, France}
\date{\today}

\begin{abstract}
NP-hard problems are not believed to be exactly solvable through general polynomial time algorithms. Hybrid quantum-classical algorithms to address such combinatorial problems have been of great interest in the past few years. Such algorithms are heuristic in nature and aim to obtain an approximate solution. Significant improvements in computational time and/or the ability to treat large problems are some of the principal promises of quantum computing in this regard. The hardware, however, is still in its infancy and the current Noisy Intermediate Scale Quantum (NISQ) computers are not able to optimize industrially relevant problems. Moreover, the storage of qubits and introduction of entanglement require extreme physical conditions.
An issue with quantum optimization algorithms such as QAOA is that they scale linearly with problem size. In this paper, we build upon a proprietary methodology which scales logarithmically with problem size – opening an avenue for treating optimization problems of unprecedented scale on gate-based quantum computers. In order to test the performance of the algorithm, we first find a way to apply it to a handful of NP-hard problems: \MAXCUT, \PARTITION, \MAXCLIQUE, \MWIS. Subsequently, these algorithms are tested on a quantum simulator with graph sizes of over a hundred nodes and on a real quantum computer up to graph sizes of 256. To our knowledge, these constitute the largest realistic combinatorial optimization problems ever run on a NISQ device, overcoming previous problem sizes by almost tenfold.
\end{abstract}

\maketitle


\section{\label{sec:level1}Introduction }\label{sec:1}

NP-hard problems are problems that do not have algorithms that can give an exact solution in polynomial time, whereas it is 'easy' to verify the solution if it is known \cite{hillar2013most,woeginger2003exact,sanchez1989determining}. While finding exact solutions to large problems is difficult, there exist many algorithms that find approximate solutions to these problems \cite{klein2010approximation,hochba1997approximation,bui1992finding,hendrickx2010matrix}. In the scope of quantum computing, a huge amount of research has been carried out on hybrid quantum-classical algorithms \cite{qaoa,qaoa2, PhysRevA.106.010101, peruzzo2014variational, moll2018quantum,stokes2020quantum,nakanishi2020sequential,cerezo2021variational,bittel2021training,lubasch2020variational,mariella2021quantum,barkoutsos2020improving,tilly2022variational}. In such algorithms, quantum circuit measurements are used in tandem with a classical optimization loop to obtain an approximate solution. 

One of the most commonly used hybrid algorithms is the Quantum Approximate Optimization Algorithm (QAOA) \cite{qaoa,fuchs2021efficient,larkin2022evaluation,herrman2021globally,zhou2020quantum}. One of the main drawbacks of the QAOA is that the number of qubits required scales linearly with problem size \cite{guerreschi2019qaoa}. This means that a graph of $n$ nodes would require an $n$-qubit quantum computer to be solved. At the moment, the largest available universal gate-based quantum computer is IBM's Osprey device, containing 433 qubits. All the qubits are not of the same quality and the larger the problem, the more likely it is to obtain noisier results due to the presence of qubits with higher error rates. Moreover, these qubits are not all-in-all connected, meaning that in case of large sized problem, numerous SWAP gates would have to be used in order to run the circuit, leading to more noise. 

It is therefore not surprising that a smaller scale quantum computer is likely to provide much better results that a larger one. In light of this, an algorithm to encode the \MAXCUT problem on a quantum computer using logarithmically fewer qubits was developed \cite{rancic}. This encoding allows us to represent much larger problems using a fairly small number of qubits. Therefore a \MAXCUT problem with a graph of $n$ nodes can be represented using only $\lceil\log_2 n \rceil $ qubits. 

Since the developed algorithm deals specifically with solving the \MAXCUT problem, a logical extension of this algorithm would be to expand the applicability of the algorithm to other problems. This can be approached in two ways, as shall be demonstrated in the following sections.

The paper is structured as follows. In section \ref{sec:2}, we describe in detail the logarithmic encoding of the \MAXCUT problem on a quantum computer. In section \ref{sec:3}, we show how this algorithm can be applied on a variety of NP-hard problems by converting them, directly or indirectly, to the \MAXCUT problem. In section \ref{sec:4}, we show how any Quadratic Unconstrained Binary Optimization Problem (QUBO) problem can be treated using the logarithmic encoding. In section \ref{sec:5}, experimental results of all the methods described in the previous sections are shown. Notably, we show quantum simulator results with instances of sizes of over a hundred nodes/objects, as well as quantum hardware (QPU) results for problem sizes up to 256.

\section{Methods}

\subsection{An Introduction to the Quantum Model of Computation}

Quantum computing \cite{neilsen,Preskill2018quantumcomputingin,Preskill2021,Feynman1982} presents a new way of doing computations by making use of fundamental properties of quantum mechanics such as superposition and entanglement. A classical bit consists of 2 possible states, $0$ and $1$. In quantum computing the first building block is the quantum bit or qubit. Much like a classical bit, a qubit has 2 basis states $\ket{0}$ and $\ket{1}$.

However unlike a classical bit, a qubit can exist in any linear combination of $\ket{0}$ and $\ket{1}$. Let $\ket{\psi}$ define the state of a qubit, then it can be mathematically defined as:
\begin{equation}\label{qubit}
    \ket{\psi}=\alpha \ket{0}+\beta \ket{1}
\end{equation}

where $\alpha$ and $\beta$ are complex coefficients such that $|\alpha|^2+|\beta|^2=1$. The coefficients $\alpha$ and $\beta$ are such that $|\alpha|^2$ and $|\beta|^2$ represent the probability of a qubit being in the state $\ket{0}$ and $\ket{1}$ respectively. The existence of qubits in all the intermediate states between $0$ and $1$ is called superposition.

Another important concept is that of entanglement. Two qubits are said to be entangled if they cannot be represented as a product state. This means that they can only be written in a combined superposition of all possible two-qubit basis states. Let $\ket{\Psi}$ be an $N$-qubit state. It can be represented mathematically as follows:
\begin{equation}
    \ket{\Psi}=\sum\limits_{i\in \{0,1\}^{N} } c_i\ket{i} 
\end{equation}

To completely describe the $N$-qubit state, we will require $2^N$ coefficients. If there is no entanglement then each qubit can be expressed separately. From equation \eqref{qubit} it can be seen that we need 2 coefficients to encode a single qubit and therefore we would need only $2N$ coefficients to represent $N$ non-entangled qubits. Entanglement is therefore of paramount importance in quantum computing as it helps encode much more information.

The specific model of quantum computation used in this paper is called gate-based quantum computing. There exist other types of quantum computing such as quantum annealing \cite{hauke2020perspectives}, topological quantum computing \cite{stern2013topological} and measurement-based quantum computing \cite{briegel2009measurement}. They are, however, beyond the scope of this paper. In the gate-based model we create a quantum circuit which is a combination of qubits and operations on qubits.

Operations on qubits are known as quantum gates. Mathematically, these can be represented by square matrices. A important property of gates is that they must be Hermitian. There are 2 basic types of gates:

\begin{compactenum}
    \item Single qubit gates : They act on a single qubit and alter the state of the qubit.
    
     \item Multiple qubit gates : These gates act on multiple qubits. These are also referred to as entangling gates. The simplest entangling gate is the CNOT gate. 
\end{compactenum}
    
All quantum operations can be represented using single qubit gates and the CNOT gate. The gate-based model is therefore a universal model of quantum computation.

Once we have created a circuit out of qubits and gates we finally need to 'measure' our qubits, which is equivalent to calculating the expectation value the system.  Let all the gates in the circuit be represented by unitary matrix $U$. If it is applied to the initial state $\ket{\Psi}$, we have the following final state:

\begin{equation}
    \ket{\Psi'}=U\ket{\Psi}
\end{equation}

The expectation value of a measurable $H$ is given by the following expression:
\begin{equation}
    \braket{H}=\bra{\Psi'}H\ket{\Psi'}
\end{equation}

A measurable can be mathematically represented by a Hermitian matrix.

Since quantum computers were first theorized in the 1980s, theoretical advantages of quantum algorithms compared to their classical counterparts have been proven for several problems \cite{shor1999polynomial, grover1996fast,kerenidis2020quantum}. However, implementing them requires a significant amount of high quality quantum resources. Specifically, they need quantum computers with many qubits as well as the ability to handle a lot of quantum operations (gates) without generating much noise. In other words, they should be able to run deep quantum circuits on several qubits. Such quantum computers might become available in the medium term but at present we have noisy intermediate-scale quantum (NISQ) computers, which cannot handle these algorithms. This has led to a significant amount of research in the development of hybrid algorithms that can run on currently available hardware.

\color{black}

\subsection{A qubit-efficient
\MAXCUT Algorithm}\label{sec:2}

Contemporary quantum optimization algorithms in general scale linearly with problem size. This means that if the problem consists of an $n$ node graph, the algorithm will require $n$ qubits to solve the problem. Note that to solve a problem here implies to obtain an approximate solution. Following Ref. \cite{rancic}, we present an algorithm that scales logarithmically with the problem size. For a problem of size $n$, the number of qubits required is $\lceil \log_2 n \rceil$.

\subsubsection{Description of the algorithm}

Recall first the definition of \MAXCUT:

\probdefbis%
{A weighted graph $G(V,E,w)$.}
{Find $x \in \{1,-1\}^{|V|}$ that maximizes $\sum_{ij} w_{ij} \dfrac{(1-x_i)(1+x_j)}{4} \forall \{(i,j)\in E\}$, where $w_{ij}$ are the weights on the edges.}
{\MAXCUT}{}{def:MAXCUT}

Given a graph $G(V,E)$, the \MAXCUT can be represented using the graph Laplacian matrix. The graph Laplacian is defined as follows:

\begin{equation}\label{laplacian}
L_{ij}=\begin{cases}
          degree(i) \quad &\text{if } \, i=j  \\
          -w_{ij} \quad &\text{if } \, i \neq j \text{ and } (i,j) \in E \\
          0 \quad &\text{otherwise}
     \end{cases}
\end{equation}

Note that the degree here is a weighted degree.

The \MAXCUT value is given by the following equation \cite{pothen}: 
\begin{equation}\label{eqn1}
\text{\MAXCUT}=\frac{1}{4}x^T L x
\end{equation}
where $L$ is the Laplacian matrix and $x \in \{1,-1\}^{|V|}$ is the bi-partition vector. 

Due to fact that the Laplacian is a Hermitian matrix, it resembles a Hamiltonian of an actual physical system. The quantum analog of equation \eqref{eqn1} is
\begin{equation}\label{eqn2}
    C(\theta_1... \theta_n)=2^{n-2}\bra{\Psi(\theta_1...\theta_n)}L\ket{\Psi(\theta_1...\theta_n)}
\end{equation}
where $L$ is the Laplacian matrix of the graph, $\ket{\Psi}$ is the parameterized ansatz, $n$ is the size of the graph, and $\theta = \{\theta_1....\theta_n\}$ are the variables to be optimized. $2^{n-2}$ is the normalization constant. 

\begin{figure}[!htb]
\begin{center}

  \includegraphics[width=\linewidth]{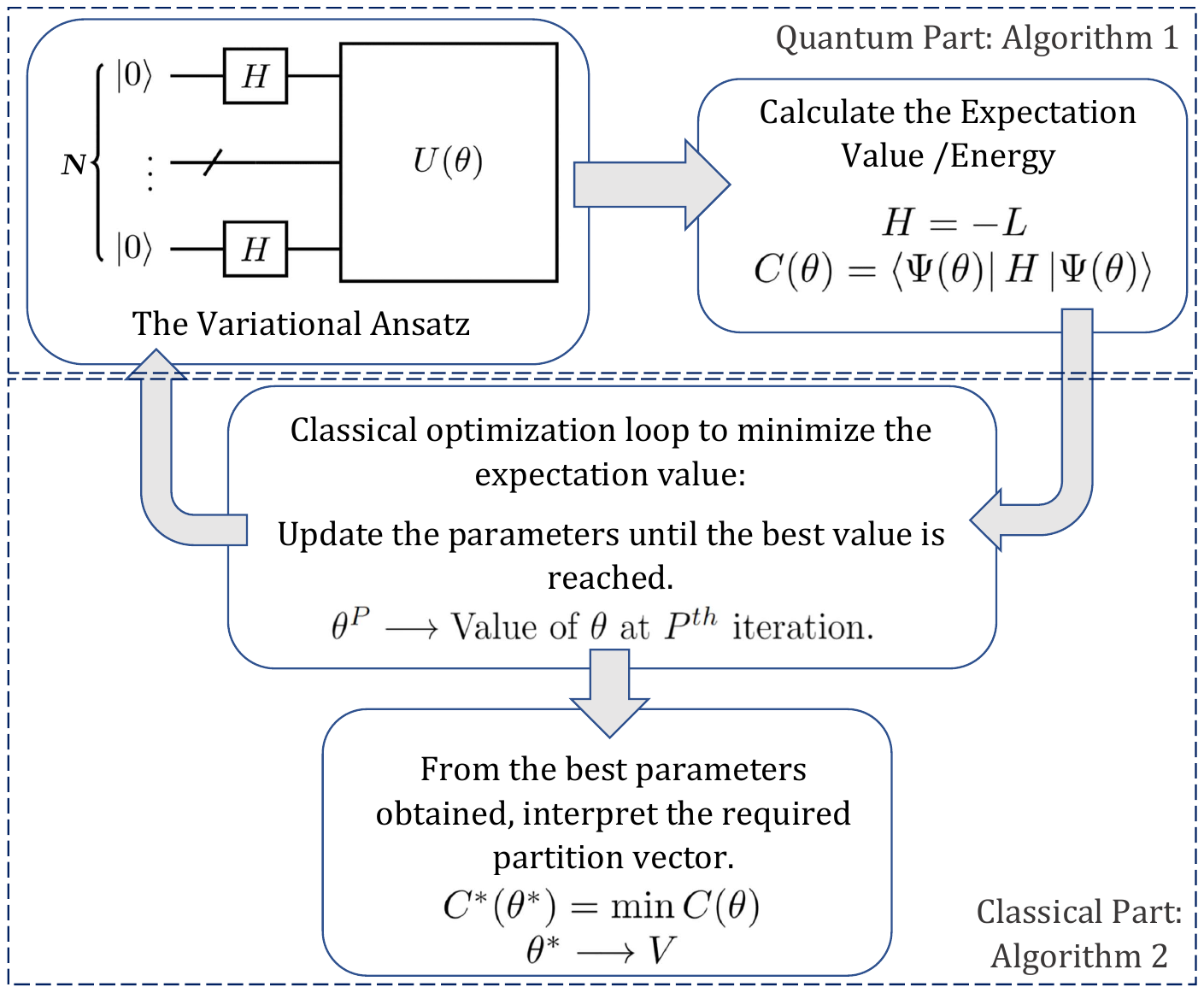}
  \caption{Diagrammatic representation of the hybrid quantum-classical algorithm.}
  \label{algprocess}
  \end{center}
\end{figure}

As described in Figure \ref{algprocess}, we have designed a variational algorithm that finds a good approximation to the best \MAXCUT. Starting from the initial values of $\theta$ parameters, we call a quantum circuit to evaluate the objective function (Algorithm \ref{alg:one}) and run a classical black box optimization loop over the $\theta$ parameters (Algorithm \ref{alg:two}). As a result, we obtain $\theta^*$ to evaluate the best solution.

To evaluate the expectation value $C$ on a quantum computer, first we need to create the ansatz $\ket{\Psi(\theta_1...\theta_n)}$. In order to do this the following steps are required.
\begin{compactenum}[1.]
    \item We define a function $R(\theta_k)$ as follows:
    \begin{equation}
        R(\theta_k)=\begin{cases}
          0\quad &\text{if } \, 0 \leq \theta_k <\pi \\
          1 \quad &\text{if } \, \pi \leq \theta_k < 2\pi \\
     \end{cases}
    \end{equation}
    Therefore,
    \begin{equation}\label{eqn5}
        \exp (i\pi R(\theta_k))=\begin{cases}
          1\quad &\text{if } \, 0 \leq \theta_k <\pi \\
          -1 \quad &\text{if } \, \pi \leq \theta_k < 2\pi \\
     \end{cases}
    \end{equation}
    
    The point of doing this is that not all classical optimizers accept binary variables. The function $R$ converts a continuous variable into a binary one, which is what we need.
    \item Given a graph $G(V,E)$ such that $|V|=n$ and $|E|=m$, to create the ansatz we first define the number of qubits required as follows:
    \begin{equation}
        N = \lceil \log_2 n \rceil
    \end{equation}
    When the number of nodes are not an exact power of 2, we can adjust $L$ to be of size $2^N$ by adding null matrices of size $2^N-|V|$, $\mathbb{O}_{2^N-|V|}$, as shown in line 3, Algorithm \ref{alg:one}.
    \item Create a quantum circuit and apply a Hadamard gate to all the qubits to achieve an equal superposition of the states (lines 7 and 8, Algorithm \ref{alg:one}).
    \item To the circuit, apply a diagonal gate $U$ (line 9, Algorithm \ref{alg:one}) of the following form:
    \begin{equation}
        U(\theta)=  \begin{bmatrix}
                e^{i \pi R(\theta_1)} & 0 & 0 & ....\\
                0 & e^{i \pi R(\theta_2)} & 0 & ....\\
                ....&....& ....&....\\
                0&0&0&e^{i \pi R(\theta_n)}
            \end{bmatrix}
    \end{equation}
    Therefore the final ansatz is:
    \begin{equation}
        \ket{\Psi(\theta)}=U(\theta)H^{\otimes N} \ket{0}^{\otimes N}
    \end{equation}
    The state in the above equation is obtained in line 10 of Algorithm \ref{alg:one}.
\end{compactenum}
    Having an ansatz, we can now define the Laplacian as an observable and evaluate the measurement (as in equation \ref{eqn2}) which is the energy of the system. Since the classical optimizer minimizes the cost function we take the negative of the Laplacian matrix. Thus the final cost function is:
    \begin{equation}\label{changedeqn2}
    C(\theta)=-2^{n-2}\bra{\Psi(\theta)}L\ket{\Psi(\theta)}
\end{equation}
    To evaluate this expectation value, the Laplacian matrix needs to be converted into a sum of tensor products of Pauli matrices (line 4 Algorithm \ref{alg:one}, see Appendix A). 
    
    Using classical black-box meta optimizers such as COBYLA, Nelder-Mead or Genetic Algorithm (as detailed in Algorithm \ref{alg:two}), we then obtain
    \begin{equation}\label{mineqn}
    C^*(\theta^*)=\min C(\theta)
\end{equation}

The final parameters obtained $\theta^*$ gives the bi-partition vector, using equation \eqref{eqn5}.

\begin{algorithm}[hbt!]
\caption{Log Encoding of \MAXCUT: Building the Objective Function}\label{alg:one}
\DontPrintSemicolon
\SetKwFunction{maxcutsolver}{EvalCost}
\SetKwFunction{classical}{ClassicalOptimizer}
\SetKwProg{Fn}{Function}{:}{}
\KwIn{Laplacian matrix of a graph $G(V,E)$}
$L \gets $Graph Laplacian of size $|V|\times |V|$\;
$N \gets \lceil \log_2 |V|\rceil$\;
$L^*\gets \begin{bmatrix}L & \mathbb{O}_{2^N-|V|} \\  \mathbb{O}_{2^N-|V|} & \mathbb{O}_{2^N-|V|}\end{bmatrix}$\;
$H \gets \dfrac{1}{n}\sum\limits_{i=1}^{4^N}Tr(J_i\cdot L^*)J_i\text{ where }J=\{\prod_{k=1}^N S^{\otimes k}  \}$\;
$\theta \gets$ List of $|V|$ parameters \;

  \Fn{\maxcutsolver{$\theta$ }}{
        $Q \gets$ Quantum Circuit of $N$ qubits\;
        Add Hadamard gate to each Qubit\;
        $U \gets$ diagonal gate $diag(\theta,R)$\;
        Apply $U$ to $Q$\;
        $F \gets ExpectationValue(Q,H)$ \;
        \KwRet $2^{|V|-2}F$\;
  }
  \;
 

\end{algorithm}

\begin{algorithm}[hbt!]
\caption{Log Encoding of \MAXCUT: Minimizing the Objective Function}\label{alg:two}
\DontPrintSemicolon
\SetKwFunction{maxcutsolver}{EvalCost}
\SetKwFunction{classical}{Optimizer}
\SetKwBlock{Repeat}{repeat}{}
\SetKw{Break}{break}
\SetKw{Continue}{continue}
\SetKwProg{Fn}{Function}{:}{}
\KwIn{\maxcutsolver{$\theta$}}


  \Fn{\classical{\maxcutsolver{$\theta$},$\theta^{\text{initial}}$ }}{
        
        \Repeat{
$\theta^p\gets \theta \text{ at }p^{th} \text{ iteration}$  \;       
$C \gets$ \maxcutsolver{$\theta^p$}\;
\eIf{C is sufficiently good}{$C^*\gets C$\;\Break} {Update $\theta^p \rightarrow \theta^{p+1}$\;\Continue}

}

        \KwRet $C^*$\;
  }
  \;
 

\end{algorithm}

\subsubsection{Advantages and disadvantages of the algorithm}

The algorithm helps us represent large problems (by current standards of quantum computing) on a quantum computer. Algorithms like the QAOA, for example, require 128 qubits to represent a 128-node \MAXCUT problem. The same problem can be solved by the proposed algorithm using only 7 qubits. A problem with over a million nodes can be represented with just 20 qubits, something that is quite unthinkable using contemporary algorithms. It therefore has the promise of being able to be applied to interesting and even industrially relevant sizes using the currently available sizes of NISQ QCs. While the QAOA depends on the availability of a large number of qubits for the algorithm to work on any interesting problem, the above algorithm only depends on an increased accuracy in QCs of current size. The number of CNOT gates required for the QAOA ansatz is $p|E|$, where $p$ is the depth of the algorithm (for all practical purposes, $p\gg 1$ \cite{harrigan2021quantum}) and $|E|$ is the number of edges in the graph. In our algorithm the number of CNOTs is equal to $|V|-1$, $|V|$ being the number of vertices. Generally, and especially at higher densities, it is easy to see that $p|E|\gg |V|$. This means that our circuit turns out to be much shallower than that of QAOA.

In our study, the search space remains the same as that of the classical search space. A procedure to reduce the number of variables has been presented in \cite{rancic}. However, this method is beyond the scope of the current work. Readers should also refer to followup studies \cite{winderl2022comparative} where the algorithm was evaluated on the \MAXCUT problem by using the alternating optimization procedure \cite{bezdek2002some} which scales polynomially in problem size.

\subsection{Applying the algorithm to other NP-hard problems}\label{sec:3}
\begin{figure}[htb!]
\begin{center}
  \includegraphics[width=\linewidth]{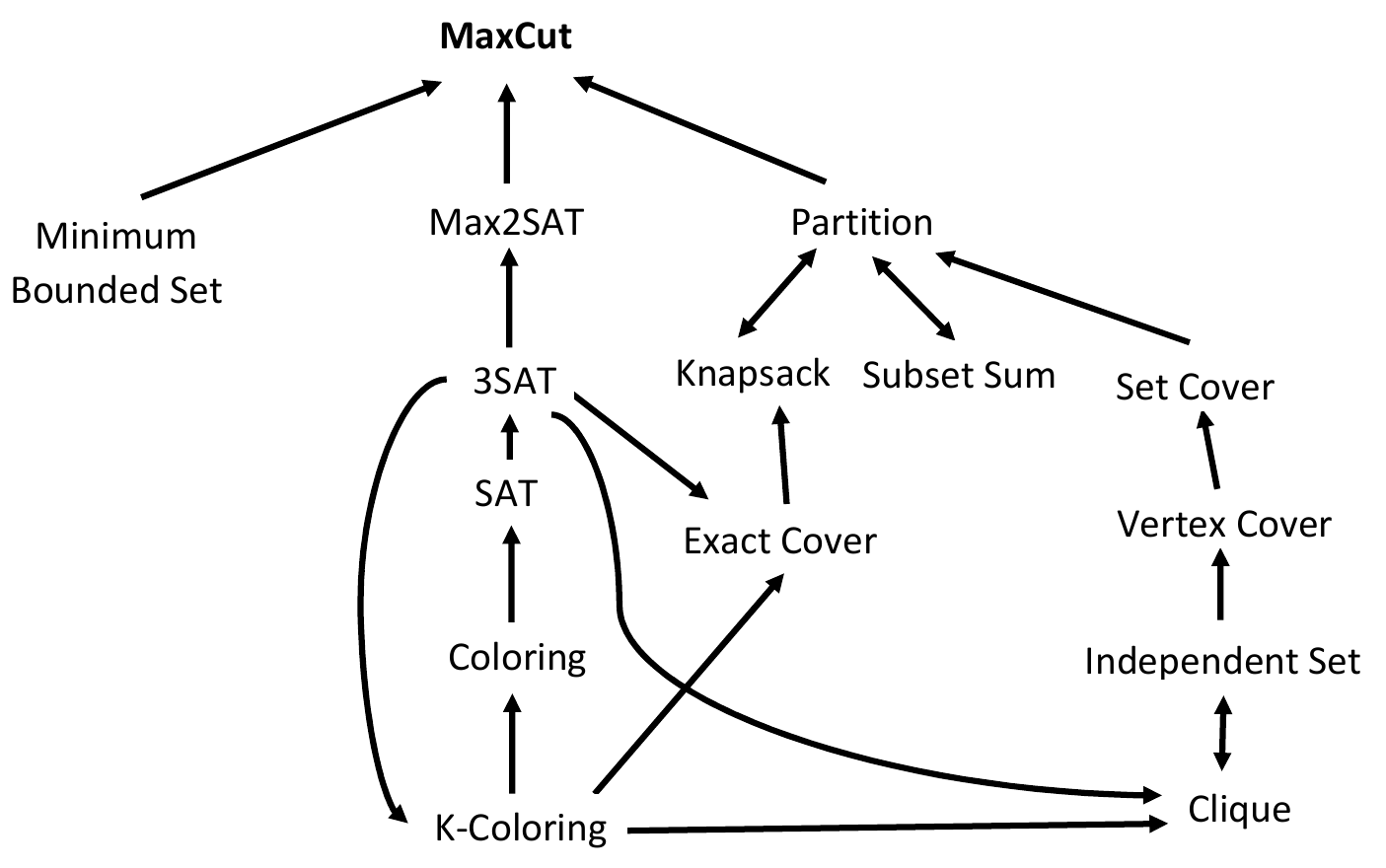}
  \caption{Graph of \MAXCUT transformation family for NP-complete problems.}
  \label{conv}
  \end{center}
\end{figure}

A logical next step is to attempt to solve a variety of combinatorial optimisation problems using the algorithm. In Karp's paper from 1972 \cite{karp}, he outlined how we can convert one NP-complete problem into another. A more recent paper \cite{surveypoly} lists numerous more such reductions. Figure \ref{conv} shows a subset (a transformation family) of these reductions directly or indirectly relating to \MAXCUT. Here, we follow a similar logic to convert various NP-hard problems to \MAXCUT. 

Note, however, that these conversions might not have a one-to-one scaling. 
For example, an $n$ variable \MAXSSAT problem requires us to solve a $2n$ node  \MAXCUT problem.

In Karp's paper all the transitions are from one decision problem to another. Usually in classical computing it would be considered trivial to convert a decision problem into an optimisation one. However, our algorithm is inherently an optimisation algorithm and moreover will give various results for a various runs. The point being, it will not respond well to yes-no decision problems. Therefore, it is important to make reductions between the optimisation versions. Moreover it is important to make sure that these conversions support a wide definition of the problems (for example \MAXSSSAT instead of \SSSAT).

Following are some such polynomial-time reductions of NP-hard problems:

\subsubsection{\PARTITION to \MAXCUT}\label{ParttoMax}

\probdefbis%
{A set $S=\{w:w\in \mathbb{Z}^+\}$.}
{Find $A\subseteq S$ that minimizes \\ $|\sum\limits_{w_k\in A }{w_k}-\sum\limits_{w_l\notin A}{w_l}| $.}
{\PARTITION}{}{def:PARTITION}

This can be converted to the maximum cut problem in the following manner:
\begin{compactenum}[1.]
    \item Create a graph such that there is a node for every number.
    \item For every pair of nodes $(i,j)$, connect them using an edge of weight $w_i*w_j$.
    \item The maximum cut value of this graph gives a bi-partition that is equivalent to the minimum partition.
\end{compactenum}

\subsubsection{\MAXSSAT to \MAXCUT} \label{2satmax}

\probdefbis%
{A set of $m$ clauses $C=\{ w_{pq}(x_p+x_q) : x_p,x_q \in X \cup X'\}$ where $X=\{x:x\in \{0,1\}\}$, $X'=\{\overline{x}:x\in X\}$ and $w_{pq}$ are the clause weights. }
{Find the variable assignment $X$ that maximizes the combined weight of the satisfied clauses.}
{\MAXSSAT}{}{def:MAXSSAT}

The problem is said to be satisfiable if all the clauses are satisfied.

We can convert this problem into the maximum cut problem in the following manner:
\begin{compactenum}[1.]
    \item In a graph, assign 2 nodes for every variable, one for the variable and another for the complement of the variable. Hence there are $2|X|$ nodes in the graph.
    \item Draw an edge between the nodes representing the variables and their complements. For example connect $x_1$ and $\overline{x_1}$, $x_2$ and $\overline{x_2}$ and so on. Add a large edge weight (about $10^m$). This is to make sure that the variables and their complements do not fall in the same partition.
    \item For every clause, add an edge between the respective nodes with edge weight as $w_{pq}$. 
    \item The maximum cut of this graph is equivalent to the \MAXSSAT solution. 
\end{compactenum}

\subsubsection{\MAXCLIQUE to \MAXSSAT} \label{clqtosat}

\probdefbis%
{A graph $G(V,E)$.}
{Maximize $|V'|$ in the graph $ \{ G'(V',E') : V' \subseteq V,E'\subseteq E ,|E'| = \frac{|V'|(|V'|-1)}{2}\} $.}
{\MAXCLIQUE}{}{def:MAXCLIQUE}

It can be converted to the \MAXSSAT problem in the following manner:  
\begin{compactenum}[1.]
    \item Consider a graph $G(V,E)$ having vertices $v_i\in V$. For each vertex $v_i$ add a variable $x_i$. Also an auxiliary variable $z$. We therefore have $|V|+1$ variables.
    \item For every variable add the following 2 clauses: $(x_i+z)$ and $(x_i+\overline{z})$. Let us refer to these clauses as Type A clauses. 
    
    \item Add the following clauses: 
    $(\overline{x_i}+\overline{x_j})\text{ } \forall \text{ }(i,j)\notin E$. We will refer to these clauses as clauses of type B.
    \item The clauses of type A ensure that the maximum number of nodes are selected and the clauses of type B make sure that the selected subgraph is a clique. 
    \item To the type B clauses, add a large weight. Due to the nature of the algorithm and it's susceptibility to errors, we may get solutions that are not cliques at all. Moreover finding a clique and maximizing it are 2 different problems and by adding weights we make sure that they are not affected by one another.
    \item The \MAXSSAT problem is solved for this set of clauses. The partition of selected variables form the \MAXCLIQUE.
\end{compactenum}

\subsection{Generalizing the Algorithm}\label{sec:4}

The algorithm described above solves, originally, the \MAXCUT problem. Various conversions are then used in order to solve other problems. Here, a second, more general approach, shall be described, where any problem which can be written in the form of a Quadratic Unconstrained Binary Optimization (QUBO) problem \cite{qubotutorial} can be solved. Instead of taking the Laplacian matrix as the input, this algorithm takes as input the QUBO matrix of the problem.

Firstly, we define the QUBO matrix. To describe a problem as a QUBO, all the terms in the objective function should be either linear or quadratic. Since the variables in the objective function are binary, a linear term can be easily converted to a quadratic one, since $x_i^2=x_i \text{ } \forall\text{ } x\in \{0,1\}$.

Consider the objective function of the following form:

\begin{equation}\label{quboform}
P =  \sum_{\substack{ij}} a_{ij} x_i x_j
\end{equation}

It can be rewritten as:
\begin{equation}\label{quboeqn}
    P= \begin{pmatrix}
          x_1 & x_2 & ... & x_n
    \end{pmatrix}
    \begin{pmatrix}
          a_{11} & a_{12} & ... & a_{1n} \\
          a_{21} & a_{22} & ... & a_{2n} \\
          ... & ... & ... & ... \\
          a_{n1} & a_{n2} & ... & a_{nn}
    \end{pmatrix}\begin{pmatrix}
          x_1 \\ x_2 \\ ... \\x_n
    \end{pmatrix}
\end{equation}
\begin{equation}
    P= x^T
    Qx
\end{equation}

\begin{equation}\label{eq:Q}
    Q= \begin{pmatrix}
          a_{11} & a_{12} & ... & a_{1n} \\
          a_{21} & a_{22} & ... & a_{2n} \\
          ... & ... & ... & ... \\
          a_{n1} & a_{n2} & ... & a_{nn}
    \end{pmatrix}
\end{equation}

$Q$ is the required QUBO matrix.

This matrix cannot be directly used in the algorithm. This is because in the original \MAXCUT algorithm, the variable used belongs to the set $\{1,-1\}$ and not $\{0,1\}$. To make the equation mathematically consistent, we need to reformulate the QUBO matrix. 

Let $z\in \{1,-1\}$ and $x\in \{0,1\}$, then $x=\dfrac{1-z}{2}$. Equation \eqref{quboform} therefore becomes:

\begin{equation}\label{qubonew}
P = \sum_{i} a_{ii} \dfrac{1-z_i}{2} + \sum_{\substack{ij \\ i \neq j}} a_{ij} \dfrac{1-z_i}{2} \dfrac{1-z_j}{2}
\end{equation}

Note that in the first term $\dfrac{1-z_i}{2}$ has been used instead of $\bigg(\dfrac{1-z_i}{2}\bigg)^2$ since $x_i^2=x_i$. 

In the search for optimal values of parameters $z$ we can eliminate the constant terms in $P$ as they only add a constant shift to the cost function. We can therefore simplify \eqref{qubonew} as follows:

\begin{align}
P&=\sum_{i} a_{ii} \dfrac{1-z_i}{2} + \sum_{\substack{ij \\ i \neq j}} a_{ij} \dfrac{1-z_i}{2} \dfrac{1-z_j}{2}\\
       &=\dfrac{1}{2}\sum_{i} a_{ii} (1-z_i) + \dfrac{1}{4}\sum_{\substack{ij \\ i \neq j}} a_{ij} (1-z_i-z_j+z_iz_j)\\
       &=-\dfrac{1}{2}\sum_{i} a_{ii} z_i + \dfrac{1}{4}\sum_{\substack{ij \\ i \neq j}} a_{ij} (-z_i-z_j+z_iz_j)\label{qubonew1} 
\end{align}

The above \textit{cannot} be represented in a matrix form similar to Eq. \ref{eq:Q} since it has linear terms that cannot be quadratized since $z_i^2 \neq z_i$.

We therefore need to reformulate the problem. In this reformulation the linear terms are represented in the off-diagonal terms instead of the diagonals. Let us have $2n$ variables $\{z_1,z_2....z_{2n}\}$ where $z_1... z_n \in \{1,-1\}$ and $z_{n+1}... z_{2n} \in \{1\}$. Then to represent a linear variable $z_i$, we can have the term $z_iz_{i+n}$ where $z_{i+n}=1$. The equation \eqref{qubonew1} can be rewritten as follows:

\begin{equation}\label{qubonew2}
    P=-\dfrac{1}{2}\sum_{i} a_{ii} z_iz_{i+n} + \dfrac{1}{4}\sum_{\substack{ij \\ i \neq j}} a_{ij} (-z_iz_{i+n}-z_jz_{j+n}+z_iz_j)
\end{equation}

This is our reformulated QUBO, which we shall call the \textit{spin-QUBO}(sQUBO). This is a  matrix of size $2n\times 2n$. It will require $\lceil \log_2 2n\rceil = (1+\lceil \log_2 n\rceil$) qubits, or, in other words, 1 more qubit than the original algorithm. Note that this is still an optimization problem of $n$ variables since the variables $z_{n+1}... z_{2n}$ are fixed.

Using our formulation, we propose that any problem that can be represented in a QUBO format can be solved using the algorithm described in Algorithm \ref{alg:three}.

\begin{algorithm}[hbt!]
\caption{Log Encoding of a QUBO problem: Building the Objective Function}\label{alg:three}
\DontPrintSemicolon
\SetKwFunction{maxcutsolver}{EvalCost}
\SetKwFunction{classical}{ClassicalOptimizer}
\SetKwProg{Fn}{Function}{:}{}
\KwIn{QUBO Matrix}
Convert $QUBO$ to $sQUBO$ \;
$Q \gets $sQUBO$ $\;
$N \gets \lceil \log_2 2n\rceil  $\;
$Q^*\gets \begin{bmatrix}Q & \mathbb{O}_{2^N-n} \\  \mathbb{O}_{2^N-n} & \mathbb{O}_{2^N-n}\end{bmatrix}$\;
$H \gets \dfrac{1}{n}\sum\limits_{i=1}^{4^N}Tr(J_i\cdot Q^*)J_i\text{ where }J=\{\prod_{k=1}^N S^{\otimes k}  \}$\;
$\theta \gets$ List of $n$ parameters \;

  \Fn{\maxcutsolver{$\theta$ }}{
        $QC \gets$ Quantum Circuit of N qubits\;
        Add Hadamard gate to each Qubit\;
        $U \gets$ diagonal gate $diag(\theta,R)$\;
        Apply $U$ to $QC$\;
        $F \gets ExpectationValue(Q,H)$ \;
        \KwRet $F$\;
  }
  \;
 

\end{algorithm}

\subsubsection{\MWIS using QUBO}\label{mwisqubo}

\probdefbis%
{A graph $G(V,E)$ with node weights $w_i$}
{Find $x\in \{0,1\}^{|V|}$ that maximizes $\sum_{i} w_{i}x_{i}$ such that   ${x_i+x_j\leq 1}\text{ }\forall\text{ } (i,j)\in E$.}
{\MWIS}{}{def:MWIS}
The \MWIS problem consists of an objective function and constraints. We can however incorporate the constraints in the objective function as penalty terms. Let $p$ be the magnitude of the penalty.

\begin{equation}
    W=\max \Big(-\sum\limits_{i}w_ix_i +p\Big(\sum\limits_{(i,j)\in E}x_ix_j\Big)\Big)
\end{equation}

Since $x_i$ is binary

\begin{equation}
    W=\max \Big(-\sum\limits_{i}w_ix_i^2 +p\Big(\sum\limits_{(i,j)\in E}x_ix_j\Big)\Big)
\end{equation}

Hence we have a QUBO matrix of the following form:
\begin{equation}
Q_{ij}=\begin{cases}
          -w_i \quad &\text{if } \,(i,j)\in E \text{ and } i=j  \\
          \frac{p}{2} \quad &\text{if } \,(i,j)\in E \text{ and } i \neq j \\
          0 \quad &\text{if } \, (i,j)\notin E  \\
     \end{cases}
\end{equation}

$Q_{ij}$ can now be used as input in Algorithm \ref{alg:three} to solve the problem.

\section{Results and Discussions}\label{sec:5}

In this section we first show the performance of the algorithm for the \MAXCUT problem. We compare the results from our algorithm with the optimal solution achieved using an integer linear program. We test our algorithm on both a quantum simulator and real hardware. Then, the effect of increasing graph density on performance is tested to surpass the sparse examples found in the literature. Finally for \MAXCUT, quantum simulator runs of up to 256 nodes are shown. Then we display the results of the \PARTITION problem, which has been solved by converting it to the \MAXCUT problem.

Next, the results from the QUBO method are shown. The \PARTITION problem is solved, this time using the QUBO method, and the results are compared with the conversion method.

\subsection{\MAXCUT}

We start by benchmarking the \MAXCUT algorithm against classical methods such as 0-1 integer linear programming and Goemans-Williamson method. All graph instances in this section are generated using the $fast\_gnp\_random\_graph()$ function of the networkx Python package, with $seed = 0$ for all cases.  
\begin{figure*}[ht]
	\begin{minipage}{\textwidth}
	\begin{overpic}[width=0.5\textwidth]{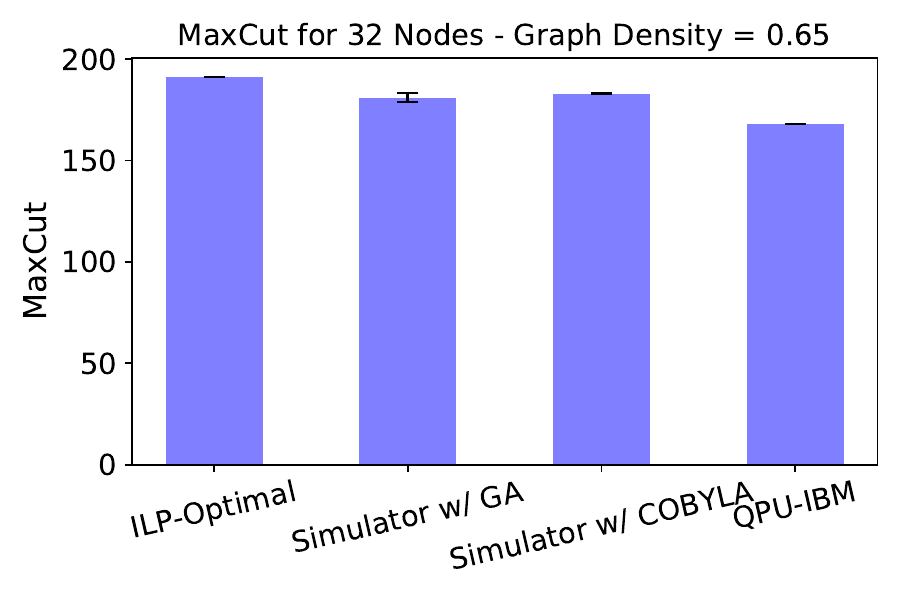}
	\put(0,0){\textbf{a)}}
	\end{overpic}\begin{overpic}[width=0.5\textwidth]{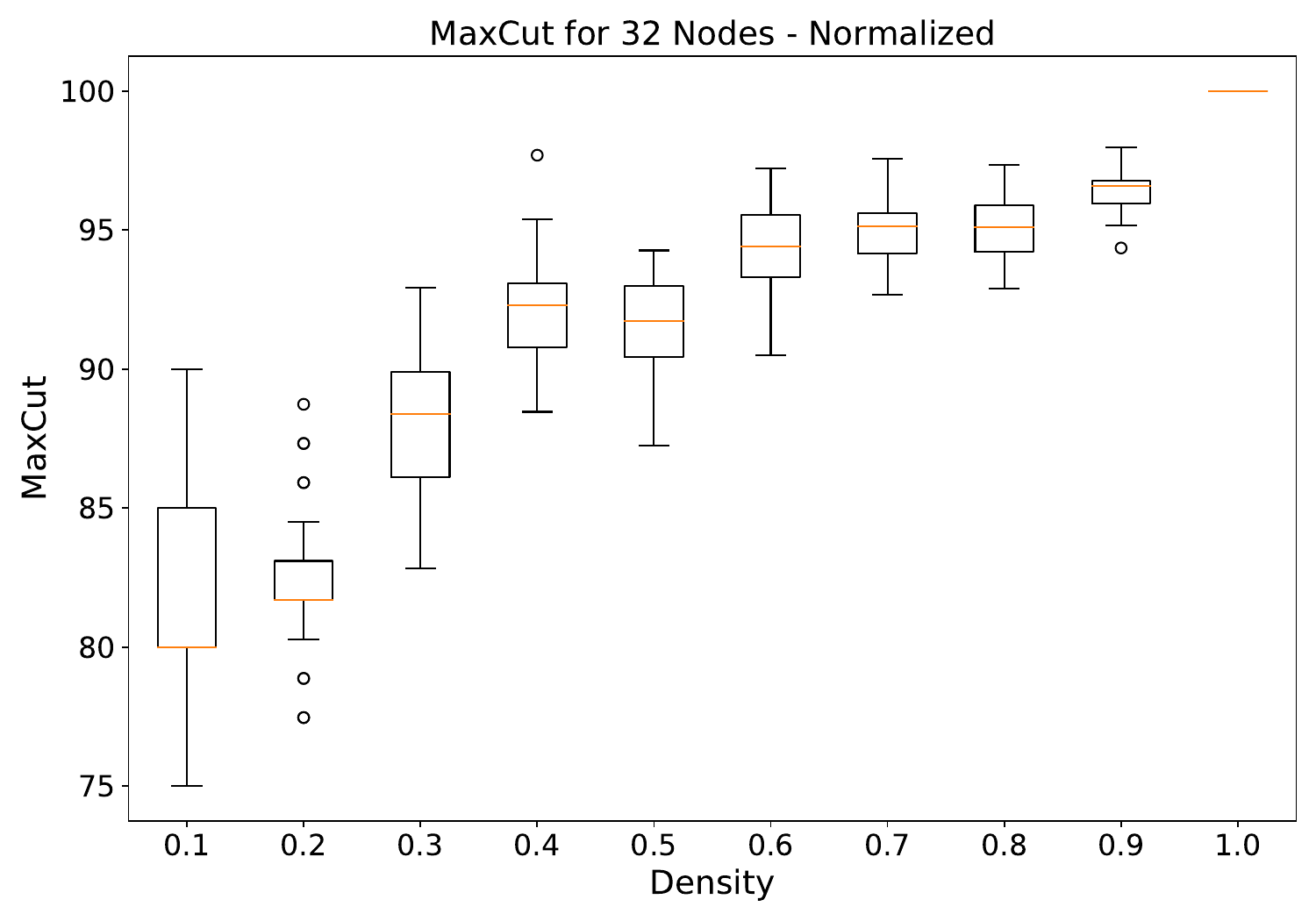}
	\put(0,0){\textbf{b)}}
	\end{overpic}
	\end{minipage}
 \caption{\textbf{a)} Performance of the algorithm on a 32-Node graph instance. The QPU result is based on a single run while the simulator results are based on 50 runs. \textbf{b)} The \MAXCUT of 32-node graphs of varying densities. Optimizer used is Genetic Algorithm. Data is based on 50 runs for each instance and is normalized using the optimal result obtained using ILP.\label{maxcut1}}
 \end{figure*}

Figure \ref{maxcut1}a) shows the performance of the algorithm versus the optimal solution obtained using an Integer Linear Program (see Appendix B). Two different classical optimizers have been used for the runs on the Quantum Simulator. We can see that both classical optimizers give fairly similar results, $90.04\% $ of the optimal for the genetic algorithm and $91.04\%$ of the optimal for COBYLA. The result from the QPU is slightly worse ($83.58\%$ of the optimal), as is expected due to the noise present in the current devices. Note that only a single instance has been considered here as opposed to multiple. This is because running algorithms on real hardware is extremely time consuming due to queue times (wait times).

In Figure \ref{maxcut1}b), 10 randomly generated 32-Node instances are tested with increasing graph density. Here graph density implies the fraction of the total possible edges present in the graph. For each instance, data was collected for 50 runs, using 2 different types of classical optimizers, COBYLA and Genetic Algorithm. In addition, a 0-1 integer linear program (ILP) \cite{maxcutmilp} was used to obtain the optimal result of each of the instances.  The ILP data is then used to normalize the simulator data. Hence, the data is in the form of percentage of optimal value. In most instances, the genetic algorithm (GA) performed better than COBYLA. This, however, might be down to the fact that the genetic algorithm is simply able to cover a larger search space. Larger instances might require a larger number of iterations with high computational cost. In these cases, using COBYLA could be more practical. While the GA results vary with each run, the results from COBYLA are the same in each run. This can be seen from the fact that the COBYLA plot has a flat error bar. We can see that the increase in the number edges does not heavily impact the accuracy of the algorithm. This is an important factor and is useful for the sections to follow.

\begin{table*}
\begin{center}
  \begin{tabular}{|l|l|l|l|l|l|l|l|}
    \hline
    \multirow{2}{*}{Graph Density} &
      \multicolumn{3}{c|}{ILP} &
      \multicolumn{2}{c|}{Quantum Simulator} &
      \multicolumn{2}{c|}{QPU} \\ \cline{2-8} 
    & Solution & Time(s) & Gap($\%$) & Solution & Time(s) & Solution & Time(s) \\
    \hline
    0.30 & 383 & 7.6 & 3.92 & 343.9 &280 & 282 & 383 \\ 
0.35& 443 & 69.7 & 3.84 & 400.8 & 272& 365 & 439\\ 
0.40& 497 & 1077.7& 3.82 &  454.3 &  270 & 380 &  393\\ 
0.45 & 553 & 1375.2 & 3.98 & 512.8 & 272 & 446 & 518 \\
\hline
  \end{tabular}
  \caption{64-Node \MAXCUT results\label{table1}}
\end{center}
\end{table*}

\begin{table*}[htb!]
\begin{center}
  \begin{tabular}{|l|l|l|l|l|l|l|}
    \hline
    \multirow{2}{*}{Graph Density} &
      \multicolumn{3}{c|}{128 Nodes} &
      \multicolumn{3}{c|}{256 Nodes} \\ \cline{2-7} 
    & GW Range& Solution & $\%$ Difference &GW Range & Solution &  $\%$ Difference \\
    \hline
    0.3& 1376 - 1431 &1270 & 88.7 - 92.3 &5408 - 5587 & 5066 & 90.7 - 93.7 \\ 
0.4& 1796 - 1864 & 1691 & 90.7 - 94.1 & 7087 - 7232  & 6736 & 93.1 - 95.0 \\ 
0.5& 2186 - 2271 & 2103& 92.6 - 96.2 & 8701 - 8880 & 8367 & 94.2 - 96.2 \\ 
0.6 & 2618 - 2679 & 2501 & 93.3 - 95.5 & 10356 - 10504 &  9967 & 94.9 - 96.2 \\
\hline
  \end{tabular}
  \caption{128 and 256-Node \MAXCUT results using a quantum simulator. \label{table2}}
  \end{center}
\end{table*}

In order to demonstrate the scalability of the algorithm, we further test the algorithm on problem instances of 64, 128 and 256 nodes (6, 7 and 8 qubits respectively). For the case of 64 node graphs, as shown in Table \ref{table1}, each instance is run 10 times on the quantum simulator and their mean and standard deviation are shown. The genetic optimizer is used for all obtained data. The data was normalized by using the ILP algorithm as mentioned before. The model solved the problem upto a specified integrality gap of $4\%$. The data represented in the table is given as a percent of the solution obtained from ILP.  We can see that for all cases, the results are nearly or over 90\% of the optimal cut. It is seen again that increase in graph density does not affect performance whatsoever. Furthermore, the execution time of ILP increases rapidly with graph density whereas it remains more or less the same for both the simulator and the QPU. For example in the graph with density $0.45$, the ILP takes $1375.2$ seconds whereas the quantum simulator and QPU take $272$ seconds and $518$ seconds respectively.

For 128 and 256 nodes (Table \ref{table2} and \ref{qpumaxcut}), the Goemans Williamson (GW) method \cite{goemans1995improved} is used for benchmarking. This is because the ILP took longer than 2 days without converging (Gurobi optimizer) on a PC. The GW range is based on 50 runs. Table \ref{table2} shows results using a quantum simulator while Table \ref{qpumaxcut} shows results obtained using an IBM quantum computer. The $ibmq\_mumbai$ backend was used for the instances of size 128 while the $ibmq\_guadalupe$ was used for the instance of size 256.

\begin{table}[htb!]
\begin{center}
  \begin{tabular}{|c|c|c|c|}
    \hline
   Instance & Solution & GW Range & $\%$ Diff. \\ 
   \hline
   Size=128, Density=0.4 & 1538 & 1796 - 1864
 & 82.5 - 85.6 \\
   Size=128, Density=0.5 & 2022 & 2186 - 2271 & 89.0 - 92.5\\
    Size=256, Density=0.5 & 8079 & 8701 - 8880 & 90.9 - 92.8 \\
\hline
  \end{tabular}
  \caption{128 and 256-Node \MAXCUT results using QPU with GA.\label{qpumaxcut}}
\end{center}
\end{table}

\subsection{\PARTITION as a conversion from \MAXCUT}\label{Partmaxexp}

\begin{figure*}[htb!]
  \includegraphics[width=1.00\linewidth]{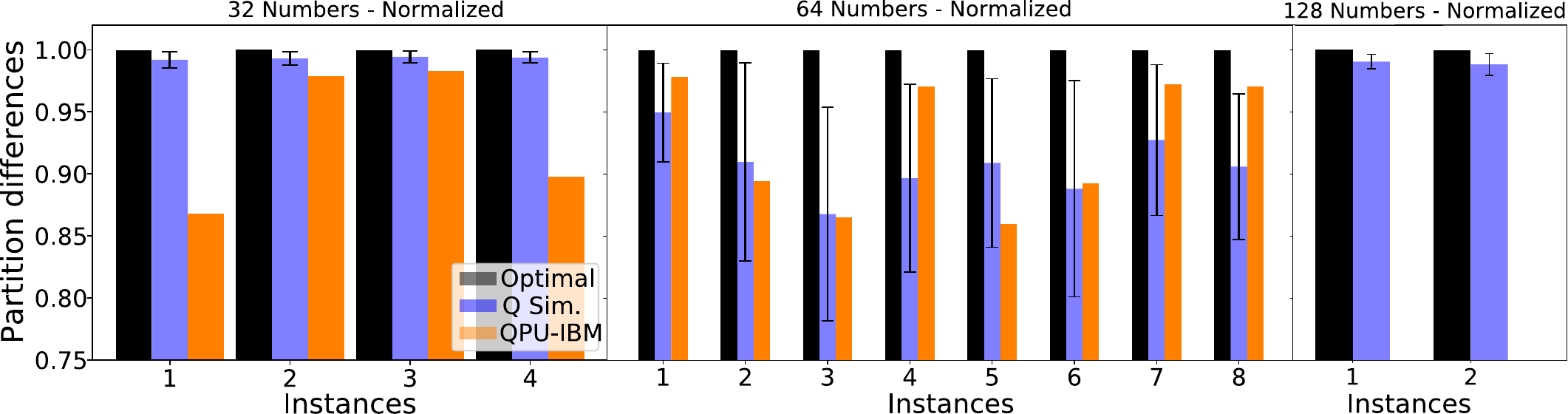}
  \caption{{The difference between partition sets for 32, 64 and 128 Numbers. Each instance was run on the quantum simulator 4 times. Each instance was run on the quantum simulator with GA, 100 times for 32 numbers, 10 times for 64 numbers and 4 times for 128 numbers. The QPU data are based on a single run. \label{partfig}}}
\end{figure*}

As described in Section \ref{ParttoMax}, the number partitioning problem can be directly converted into the \MAXCUT problem. The graphs hence formed are weighted fully-dense graphs.

For the instances, all the numbers used were random integers between 1 and 100. Tests were carried out on the quantum simulator as well as on real hardware from IBM.  

The results of partition differences have been normalized in the following manner. For a problem with $N$ numbers, if the partition difference is $p$, then the normalised difference is $p_{norm}=\dfrac{50N-p}{50N}$. All our numbers are random integers between $1$ and $100$, hence $50.5$ on an average. For simplicity we use $50$ in $p_{norm}$.

The optimal value for each instance is obtained using the Integer Quadratic model described in Appendix C.

Figure \ref{partfig} displays the performance of 32, 64, and 128-number \PARTITION converted to \MAXCUT. For 32 integers, we have a mean value of $98\%$ and a very small dispersion using the quantum simulator. For 64-number \PARTITION, the mean values are better than $85\%$ for all instances. Moreover, for both problems sizes,  despite the fact that the \PARTITION problem leads to a complete graph \MAXCUT problem, actual QPUs are able to demonstrate an approximate solution. The problem of size 128 has mean values of about $97\%$ on a quantum simulator. 

All QPU runs in this section are done on $ibmq\_mumbai$.

\subsection{\MAXCLIQUE as a conversion from \MAXCUT}

The \MAXCLIQUE problem can be converted to the \MAXCUT problem by first converting it to the \MAXSSAT (\ref{2satmax}) and then from the \MAXSSAT to \MAXCUT(\ref{clqtosat}).

After the conversion, a $n$-node \MAXCLIQUE problem requires the solution of a $2(n+1)$-node \MAXCUT. Table \ref{cliquetable} shows results for various instances run on a quantum simulator. It is seen that in half of the instances, the best solution is the optimal solution as obtained using numpy. It should be noted that our approach finds a dense subgraph and then removes the nodes with lowest degree iteratively.

\begin{table*}[htb!]
  \begin{tabular}{|c|c|c|c|c|c|c|}
    \hline
   Instance & No. of Qubits & No. of Runs & Best Solution & Worst Solution & Av. Solution & Opt. Solution \\ 
   \hline
   Size=31, Density=0.3 & 6 & 50 & \textbf{4}& 3&3.38&4\\
   Size=31, Density=0.4 & 6&50 & \textbf{5}& 3&3.92&5\\
   Size=31, Density=0.5 & 6&50 & \textbf{6}& 3&4.46&6\\
   Size=31, Density=0.6 & 6&50 & 7& 4&5.34&8\\
   Size=31, Density=0.7 & 6&50 & 8& 5&6.26&9\\
  Size=63, Density=0.5 & 7&10 & \textbf{8}& 6&6.5&8\\
    Size=63, Density=0.6 & 7&10 & 8& 6&7.2&10\\
    Size=63, Density=0.7 & 7&10 & 11& 8&9.3& 12\\
   
\hline
  \end{tabular}
  \caption{\MAXCLIQUE results using quantum simulator with GA.\label{cliquetable}}
\end{table*}

\subsection{\MWIS using QUBO method}

In this section, results of the \MWIS problem solved using the QUBO method (section \ref{mwisqubo}) is presented. 

For each figure the performance of the algorithm is shown. The data is normalized using the optimal solution found using the commercial CPLEX solver.

\begin{figure*}[ht]
\begin{center}
  \includegraphics[width=\linewidth]{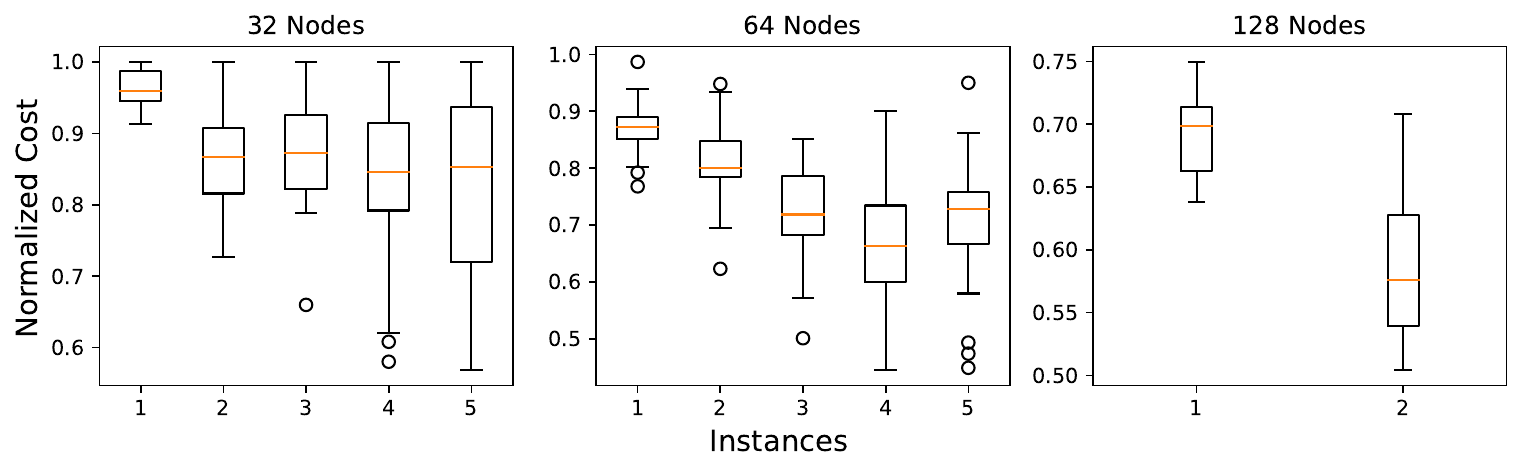}
  \caption{\MWIS problem for 32, 64 and 128-node graphs using a quantum simulator. Each instance was run on a quantum simulator with GA 50 times for graph sizes of 32 and 64 and 10 times for graph size of 128. \label{mwis}}
\end{center}
\end{figure*}

Figure \ref{mwis} shows the data for graphs of size 32, 64 and 128. The data for 32, 64 and 128 nodes is based on 50, 50 and 10 runs respectively. The data represented only takes into account the feasible solutions produced. For graphs of size 32, the mean values for all instances are above $80\%$ and the best obtained result is optimal for every instance. For graphs of size 64, the mean values for all instances are above $60\%$ and the best obtained result is on an average over $80\%$. For 128-node graphs, the solutions are slightly degraded in comparison. Note, however, that the data for 128-node instances is based only on a few runs. Moreover, the performance also depends on the number of GA iterations used in the algorithm run. The number of GA iterations used for the experiments for sizes 32, 64, and 128 are 50, 100 and 200 respectively.

Table \ref{qpumwis} shows how the performance varies depending upon the number of GA iterations used. For this table, the \MWIS Instance 4 of size 64 has been taken. An increase in the number of GA iterations not only improves the performance but also the percentage of feasible results.

\begin{table}[htb!]
\begin{center}
  \begin{tabular}{|c|c|c|c|}
    \hline
   Size & \multicolumn{1}{|p{1.6cm}|}{GA iterations} &\multicolumn{1}{|p{2.7cm}|}{Solution as \% of Optimum} &\multicolumn{1}{|p{2.3cm}|}{\% of Feasible Solutions}\\ 
   \hline
   64 &  50 & 54.8 & 60\\
   64 & 100 & 66.6 & 96\\
   64  & 200 & 77.8 & 100\\

\hline
  \end{tabular}
  \caption{\MWIS results demonstrating the relationship between GA iterations and performance. The Instance 4 of size 64 is used for this table. The performances are an average over 10 runs.\label{qpumwis}}
\end{center}
\end{table}

\subsection{A comparative study of time taken by the simulator and the QPU}
 
\begin{figure}[htb!]
\begin{center}
  \includegraphics[width=\linewidth]{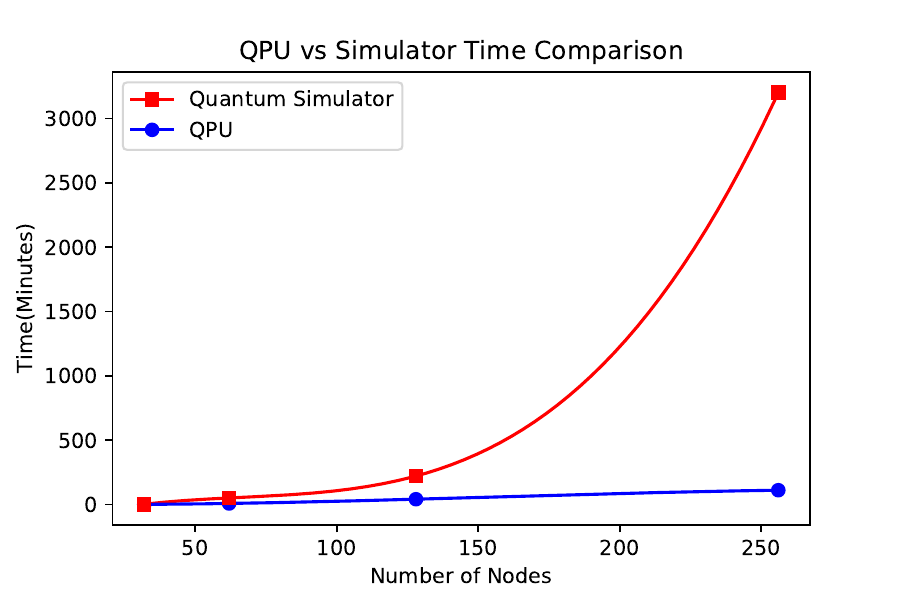}
  \caption{Plot demonstrating the time taken by the quantum simulator versus the time taken to solve the same instance on real hardware }
  \label{time}
  \end{center}
\end{figure}

 In Table \ref{table3} and Figure \ref{time}, the time taken to run the algorithm for different \MAXCUT instance sizes is compared. While the quantum computer still takes a significant amount of time to solve the problem, the time taken does not increase exponentially as in the case of the simulator. As we move towards larger instances, we reach a point where it is quicker to run a problem on a QPU than using a simulator. 
 
 Note that the QPU time here does not take into account the queue time or waiting time for the QPU runs. The real-world time was several hours or even several days for the largest run instance. This prevented us from running larger instances on the quantum computer.

\begin{table}[ht]
\begin{center}
\begin{tabular}{ |c|c|c|c| } 
\hline
N & QPU(minutes) & Quantum Simulator(minutes) \\
\hline
32 & 1.7 & 3 \\ 
64& 9 & 52 \\ 
128& 45 & 222 \\ 
256 & 112 & 3202 \\
\hline
\end{tabular}
\caption{Data for time taken for various instance sizes in the the QPU and in the quantum simulator\label{table3}}
\end{center}
\end{table}

\section{Conclusion}

In this paper, we investigated and further developed methods to logarithmically encode combinatorial optimization problems on a quantum computer. We begin with expanding the work done in \cite{rancic}, which describes a way to logarithmically encode the \MAXCUT problem. We performed several runs of this algorithm with various instances, on the quantum simulator as well as real hardware, using different classical optimizers like COBYLA and the genetic algorithm. 

We then reformulate a number of NP-hard combinatorial optimization problems into the \MAXCUT problem, either directly or indirectly and solve it on a real quantum computer. We take the \PARTITION problem as an example and solve it by using a reduction as mentioned in section \ref{ParttoMax}. This is possible since the algorithm is largely unaffected by increasing the density of the \MAXCUT graph in question, since the \PARTITION problem converts into a weighted fully-dense graph. Some performance benchmarks of the partition problem have been presented.

We then proceed to present a more general formulation inspired from the structure of the \MAXCUT algorithm. We see that instead of using the Laplacian, we can use the QUBO matrix of a problem in order to solve it. We introduce the sQUBO representation of the QUBO matrix for it to be compatible with the algorihtm. This therefore opens up the applicability of the algorithm to a wide range of algorithms. The \MWIS problem is solved using its sQUBO matrix. 

To our knowledge, it is the first time that graph problems of such sizes (256 \MAXCUT, 64 \PARTITION) have been executed on real universal gate-based quantum computers.

\section{Declarations}
\subsection{Ethical Approval and Consent to participate}
Not applicable.

\subsection{Consent for publication}
Not applicable.

\subsection{Data availability statement}
The authors will make all data available upon reasonable requests.

\subsection{Competing interests}
A part of the methodology presented in the manuscript is protected by a provisionally patent claim ”Method for optimizing a functioning relative to a set of elements and associated computer program product” submission number EP21306155.9 submitted on 26.8.2021.

\subsection{Funding}
 Y.C. and M.R. acknowledge funding from European Union’s Horizon 2020 research and innovation programme, more specifically the
$\bra{\text{NE}}\text{AS}\ket{\text{QC}}$ project under grant agreement No. 951821.

\appendix

\section{Appendix: Calculating the Expectation value of an observable}

Given a Hamiltonian matrix $H$, we first need to convert into a sum of tensor products of Pauli strings. 

Let $H$ be a $n\times n$ Hamiltonian matrix and $S =\{I,X,Y,Z\}^n=\{S_1,S_2,S_3,S_4\}^n $ be the set of Pauli matrices. We can consider $n$ to be a power of $2$ without any loss of generality. If the size of the Hamiltonian matrix is $n'$ which is not a power of $2$, we can easily convert it to a size of $n=2^{\log_2(n')}$, which is a power of $2$. The extra space in the matrix is filled with $0$'s.

This Hamiltonian can now represented on $N=\log_2(n)$ qubits. Consider the set $J=\{S_{i_1}\otimes S_{i2}...\otimes S_{iN}| i_1,i_2....i_N \in \{0,1,2,3\}  \}$ which consists of all tensor product combinations of the Pauli matrices.

Then the Hamiltonian can decomposed as:
 
\begin{equation}
    H=\sum\limits_{i=1}^{4^N} c_iJ_i
\end{equation}
where the coefficients are:
\begin{equation}
    c_{i}=\frac{1}{n}Tr(J_i\cdot H)
\end{equation}

The Hamiltonian therefore becomes:

\begin{equation}
    H=\frac{1}{n}\sum\limits_{i=1}^{4^N}Tr(J_i\cdot H)J_i
\end{equation}

The  expectation value becomes a sum of the expectation values of all the terms.

\begin{equation}
    \bra{\Psi}H\ket{\Psi}= \frac{1}{n}\sum\limits_{i=1}^{4^N}Tr(J_i\cdot H)\bra{\Psi}J_i\ket{\Psi}
\end{equation}

\section{Appendix: Integer Linear Program for \MAXCUT Problem}

Given a graph $G(V,E)$ such that $n=|V|$, and $A_{ij}$ being the corresponding Adjacency matrix terms, we have\\
\textbf{Objective :}  $\max \sum_{1\leq i\leq j\leq n}x_{ij}A_{ij}$ \\
\textbf{Constraints :}
\begin{compactenum}[1.]
    \item $x_{ij}\leq x_{ik}+x_{kj} $
    \item  $x_{ij}+x_{ik}+x_{kj}\leq 2 $
    \item $x_{ij}\in \{0,1\}$
\end{compactenum}

\section{Appendix: Integer Quadratic Program for \PARTITION Problem}

Given a set $S=\{w:w\in\mathbb{Z}^+\}$, and $A\subseteq S$, we have \\
\textbf{Variables : }$x_{i}=\begin{cases}
          1 \quad &\text{if } \, w_i \in A  \\
          0 \quad &\text{if } \, w_i \notin A  \\
     \end{cases}$ \\
\textbf{Objective : } $\min
\big(\sum\limits_{i=1}^n{w_ix_i}-\sum\limits_{i=1}^n {w_i}(1-x_i)\big)^2 $ \\

\nocite{*}

\bibliography{apssamp}

\end{document}